\documentclass[sigconf]{acmart}

\usepackage{algorithm}
\usepackage[noend]{algpseudocode}
\usepackage{tikz}
\usepackage{pgfplots}
\usepackage{balance}
\pgfplotsset{compat=1.16}
\usepackage{booktabs}
\usepackage{multirow}
\usepackage{xspace}

\newcommand{\hy}{y} 

\newcommand{\setC}{\mathcal{C}}
\newcommand{\setI}{\mathcal{I}}
\newcommand{\setS}{\mathcal{S}}

\newcommand{\rank}{r} 
\newcommand{\ndcg}{\mathrm{NDCG}}

\newcommand{\lambdarank}{LambdaRank\xspace}

\AtBeginDocument{%
  }

\newcommand{\vvcomment}[1]{}

\newcommand{\vvalternative}[1]{}

\copyrightyear{2024}
\acmYear{2024}
\setcopyright{rightsretained}
\acmConference[CIKM '24]{Proceedings of the 33rd ACM International
Conference on Information and Knowledge Management}{October 21--25,
2024}{Boise, ID, USA}
\acmBooktitle{Proceedings of the 33rd ACM International Conference on
Information and Knowledge Management (CIKM '24), October 21--25, 2024,
Boise, ID, USA}
\acmDOI{10.1145/3627673.3679943}
\acmISBN{979-8-4007-0436-9/24/10}

\begin{document}


\title{Improved Estimation of Ranks for Learning Item Recommenders with Negative Sampling}



\author{Anushya Subbiah}
\orcid{0009-0009-6005-6050}
\affiliation{%
  \institution{Google Research}
  \city{Mountain View}
  \country{USA}
}
\email{anushyas@google.com}

\author{Steffen Rendle}
\orcid{0009-0004-6389-2509}
\affiliation{%
  \institution{Google Research}
  \city{Mountain View}
  \country{USA}
}
\email{srendle@google.com}

\author{Vikram Aggarwal}
\orcid{0009-0004-1750-6806}
\affiliation{%
  \institution{Google Research}
  \city{Mountain View}
  \country{USA}
}
\email{viki@google.com}

  
\begin{abstract}
In recommendation systems, there has been a growth in the number of recommendable items (\# of movies, music, products).
When the set of recommendable items is large, training and evaluation of item recommendation models becomes computationally expensive.
To lower this cost, it has become common to sample negative items.
However, the recommendation quality can suffer from biases introduced by traditional negative sampling mechanisms.

In this work, we demonstrate the benefits from correcting the bias introduced by sampling of negatives.
We first provide sampled batch version of the well-studied WARP and \lambdarank methods.
Then, we present how these methods can benefit from improved ranking estimates.
Finally, we evaluate the recommendation quality as a result of correcting rank estimates and demonstrate that WARP and \lambdarank can be learned efficiently with negative sampling and our proposed correction technique.
\end{abstract}


\received{3 June 2024}
\received[accepted]{16 July 2024}

\maketitle

\section{Introduction}

Recommendation systems are critical at navigating the vast item space in contemporary products.
Item recommenders select the most promising items for a user based on the user's past interactions and present the items as a ranked list.
The data to train such systems is typically \emph{implicit feedback} in the form of past positive interactions (e.g., listen or watch events, clicks, etc.)~\cite{rendle:implicit}.
Training an item recommender from implicit feedback is computationally challenging since the learning algorithm needs to distinguish the \emph{positive} items from the full catalog of all items.
Sampling negative items during training has become a standard practice to lower the costs.
Instead of comparing the positive/ relevant items to all items, they compare the relevant items to a smaller subsets of sampled items.
A key challenge in negative sampling arises from the bias that it introduces into the loss functions.
Item recommendation models are typically trained for loss functions that try to move relevant items to the top of the list.
For example, \emph{\lambdarank}~\cite{burges:overview} suggests to weight a pair of positive and negative items by the difference in their rank, and \emph{Weighted Approximate-Rank Pairwise loss}~\cite{weston:wsabie} (WARP) weights based on the rank of the positive item.
However, a subsample of items provides only limited information about where the relevant item ranks in the full catalog.
In this work, we study how the rank of an item can be more reliably estimated on a subset of items and how this information can be used for training item recommenders with negative sampling.
Our work is inspired by recent work on evaluating recommender systems with sampled metrics~\cite{krichene:sampled_metrics} where evaluation metrics are computed on a sample of negative items.
We apply their proposed method for estimating the rank within a sample to learning item recommendation models from sampled negatives and study the effect of these methods in \lambdarank and WARP.

\begin{table}[t]
    \centering
    \begin{tabular}{ll}
    \toprule
         Symbol& Description  \\
    \midrule
         $\setI$ & catalog of all items \\
         $\setC$ & recommendation context (e.g., a user) \\
         $\setS \subset \setC \times \setI$ & set of training examples \\
         $\hy : \setI \times \setC  \rightarrow \mathbb{R}$ & scoring function \\
         $\rank : \setI \times \setC  \rightarrow \{1,\ldots, |\setI| \} $ & predicted rank of an item \\
         $m$ & number of negative samples \\
         $\tilde{\setI} \subseteq \setI, |\tilde{\setI}|=m$ & sampled set of items \\
         $\tilde{\rank}: \setI \times \setC  \rightarrow \{1,\ldots, m \}$ & rank of an item in a sample \\
         $\hat{\rank}: \setI \times \setC  \rightarrow \{1,\ldots, |\setI| \}$ & estimated rank on the full catalog \\
    \bottomrule
    \end{tabular}
    \caption{Symbols}
    \label{tab:symbols}
\end{table}

\section{Related Work}
Item recommendation from implicit feedback is a widely studied task~\cite{rendle:implicit}.
We shortly summarize some alternative approaches for tackling the task of item recommendation with large item catalogs.

One direction of work is to change the sampling distribution, e.g., sampling by item popularity \cite{bengio:sampledsoftmax} or sampling items that are highly ranked for a context~\cite{rendle:bprimpr,yuan:lambdafm}.
Using a more complex sampling distribution can serve as a correction, e.g., sampling according to the softmax distribution results in an unbiased softmax loss even with small sample sizes \cite{blanc:adaptive}.
Besides the higher computational costs for sampling itself, another downside of complex, context-dependent sampling distributions is that negative samples are not shared between multiple context in a batch.
This is problematic when batches of users and negative items should be processed in parallel, e.g., utilizing modern parallel hardware.
In our work, we take a different approach and investigate a simple sampling distribution (uniform sampling) but apply corrections to the statistics that we compute from the sample.

Moreover, there exist approaches for specialized loss functions that avoid sampling altogether and that can work efficiently on the full item catalog~\cite{hu:ials,krichene:isgd}.
The computational trick that they employ works for spherical loss functions over negative items.
In our work, we focus on more complex ranking losses where efficient formulations are not available.

\section{Problem Setting}

Let $\setC$ be a set of recommendation context (e.g., a user, or a sequence).
Let $\setI$ be a set of items that can be recommended -- we call this the \emph{item catalog}.
Let $\hy : \setC \times \setI \rightarrow \mathbb{R}$ be a recommendation scoring function.
This function can be used to rank items for a given context and to show the highest scoring items to the user.
$\hy$ is typically a learn-able function that is parameterized by model parameters $\theta$.
We aim to learn $\theta$ based on a set of training examples $\setS$.

Our work focuses on the task of learning the recommender from \emph{implicit feedback}~\cite{rendle:implicit}.
In this setting, the observations are positive-only observations such as clicks, listen events, etc.
The training examples are a set of such positive interactions: $\setS \subset \setC \times \setI$.
Each training context $c$ has some observed (or \emph{positive}) items $i$ (i.e., $(c,i) \in \setS$) and all other items are called \emph{unobserved} and considered \emph{negative} items.
For a training example $(c,i) \in \setS$, we would like that the recommender scores positive items higher than negative items, i.e., we would like the learned scoring function to satisfy $\hy(c,i) > \hy(c,j)$ if $(c,i) \in \setS$ and $(c,j) \not\in \setS$.
Learning algorithms use \emph{loss functions} to guide the search for optimal model parameters $\theta$.

\begin{algorithm}[t]
  \caption{Iterative Pairwise Training}\label{algo:iterative}
  \begin{algorithmic}[1]
  \State \textbf{Input:} dataset $\setS$, learning rate $\eta$
  \State \textbf{Output:} model parameters $\theta$
  \Repeat
    \State sample $(c,i) \sim \setS$ \Comment{positive observation}
    \State sample $j \sim \setI$ with probability $q(j|c,i)$ \Comment{negative item} \label{line:draw_neg}
    \State $\theta \leftarrow \theta - \eta \, \alpha(c,i,j)\frac{\partial}{\partial \theta}l(c,i,j)$ \Comment{gradient step}
  \Until{Convergence}
  \State \textbf{return} $\theta$
  \end{algorithmic}
\end{algorithm}

\subsection{Rank Dependent Losses}

Numerous loss functions have been proposed for learning from implicit feedback~\cite{rendle:implicit}.
Some sophisticated ranking techniques such as WARP~\cite{weston:wsabie} and \lambdarank~\cite{burges:overview} depend on the position of the items in a ranked list as they try to encourage ranking relevant items at the top.
The rank of an item $i$ in an item set $\setI$ using recommender $\hy$ is defined as:
\begin{align}
    \rank(i|c) := |\{ j \in \setI: \hy(j|c) > \hy(i|c) \}|+1 .    
\end{align}
Computing the rank requires scoring all items in $\setI$ and therefore is an expensive operation.
This is especially expensive when the item catalog is large.
In this work, we investigate speeding up this step by sampling with rank correction methods.

\subsection{WARP and \lambdarank}

WARP and \lambdarank are iterative algorithms that draw one positive observation and one negative observation and apply a gradient step on this sampled pair of items.
See algorithm~\ref{algo:iterative} for a generalized algorithm~\cite{rendle:implicit}.
WARP and \lambdarank differ in (i)~how they sample the negative item (i.e., in $q$) and (ii)~how they weight the update (i.e., $\alpha$). 
We will discuss these two choices next.
For WARP, the negative item is drawn with a rejection sampling algorithm.
In particular, WARP draws candidate items uniformly until it finds a candidate $j$ such that $l(c,i,j) > 0$, where $l$ is a pairwise loss function (see eqs.~\ref{eq:hinge} and \ref{eq:logistic}).
If the rejection algorithm cannot find a candidate in a predefined number of steps, it will exit and discard the step.
For \lambdarank, negative items are drawn uniformly without any rejection step, i.e., $q(j|c,i) = \frac{1}{|\setI|-1}$.

The weighting function $\alpha$ of \lambdarank depends on the rank $\rank(i|c)$ of the positive item $i$ and the rank of the negative item $j$ and is  $\alpha(c,i,j)=|\ndcg(\rank(i|c))-\ndcg(\rank(j|c))|$.
This rank is computed over all items $\setI$.
The weighting function, $\alpha$ of WARP depends only on the rank of the positive item $i$.
However instead of computing this rank over all items, WARP estimates this rank, $\hat{\rank}$, based on how many steps the rejection algorithm was run.
The weighting function is $\alpha(c,i,j)= \sum_{r=1}^{\hat{\rank}(i|c)} \frac{1}{r}$.

WARP employs usually a pairwise hinge loss, i.e.,
\begin{align}
    l(c,i,j) = \max(0, 1 - (\hy(c,i) - \hy(c,j))) \label{eq:hinge}
\end{align}
and \lambdarank a pairwise logistic loss
\begin{align}
    l(c,i,j) = \ln\sigma(\hy(c,i) - \hy(c,j)). \label{eq:logistic}
\end{align}
We want to highlight that computing the rank is a crucial operation for \lambdarank and WARP.
For \lambdarank this is computationally expensive because it computes the rank over all items.
This becomes costly for large catalogs.
WARP computes the rank through the rejection sampling algorithm.
Again, if the item is ranked high and the item catalog is large, then this operation is costly.

\section{Batch Training with Negative Sampling}

\begin{algorithm}[t]
  \caption{Sampled Batch Training}\label{algo:batch}
  \begin{algorithmic}[1]
  \State \textbf{Input:} dataset $\setS$, learning rate $\eta$
  \State \textbf{Output:} model parameters $\theta$
  \Repeat
    \State sample $\tilde{\setS} \sim 2^\setS $ s.th. $ |\tilde{\setS}|=k$ \Comment{positive observations}
    \State sample $\tilde{\setI} \sim 2^\setI $ s.th. $ |\tilde{\setI}|=m$ \Comment{negative items}
    \State $\theta \leftarrow \theta - \eta \, \sum_{j \in \tilde{\setI}}\sum_{(c,i) \in \tilde{\setS}}\alpha(c,i,j)\frac{\partial}{\partial \theta}l(c,i,j)$ \Comment{grad. step}
  \Until{Convergence}
  \State \textbf{return} $\theta$
  \end{algorithmic}
\end{algorithm}

Modern training algorithms make use of hardware that is designed for parallel execution.
The WARP and \lambdarank algorithms discussed so far are not well suited for modern hardware as they process one item at a time.
Especially the iterative rejection sampling algorithm of WARP is not a good fit.
We will propose now a simple variant that processes several negative and positive items in parallel.
Algorithm~\ref{algo:batch} shows this generalized algorithm.
It simply samples uniformly batches of $k$ positives and $m$ negatives.
A gradient step is applied to all $k \times m$ pairs.
This operation is better suited for parallel execution.

We note that the sampled batch training algorithm samples uniformly and does not reject any negative items like the iterative WARP algorithm.
However, even in the original iterative WARP algorithm, the gradient steps would not change if the rejection algorithm would accept all items because accepting an item with $l(c,i,j)=0$ would anyway have a gradient of 0 for the hinge loss.
So the purpose of the rejection algorithm in WARP is mostly to compute the rank.
For sampled batch training, we can use other ways to estimate ranks and can therefore remove the rejection algorithm.

\subsection{Estimation of Rank}

To better deal with large item catalogs, we rank items within the sampled set $\tilde\setI$ of negative items instead of computing the rank on the full catalog which would be very costly.
We define the rank in the sample $\tilde\setI$ as
\begin{align}
    \tilde{\rank}_{\tilde\setI}(i |c) =  |\{ j \in \tilde{\setI}: \hy(c,j) > \hy(c,i) \}| +1 \label{eq:rank_subsample}
\end{align}
and refer to it as the \emph{sampled rank}.
The sampled rank from a subset $\tilde\setI \subset \setI$ is a poor estimate of the true rank, $\rank$.
For example, the maximum rank is $m$ instead of $|\setI|$.
In general, it underestimates the rank because some of the items that are ranked above $i$ might be missing from $\tilde\setI$.

We will follow the ideas from \cite{krichene:sampled_metrics} that have previously analyzed this issue and provided correction methods in the context of evaluating with sampled metrics.
The sampled rank $\tilde{\rank}$ is a random variable that depends on the sampled set of negatives.
The sampled rank counts how often a sampled item is ranked above the positive item.
If negative items are sampled uniformly, then the probability of sampling a negative item ranking above the positive item depends on the true rank of the positive item and the probability is
\begin{align}
    p(\rank(j|c) > \rank(i|c)) = \frac{\rank(i|c) -1}{|\setI| - 1} . 
\end{align}
The sampled rank (eq.~\ref{eq:rank_subsample}) repeats this process $m$ times and counts the successes.
Thus the sampled rank is binomial distributed
\begin{align}
    \tilde{\rank}(i | c) \sim B\left(m, p\right) + 1 \quad \text{with}\quad   p = \frac{\rank(i|c) - 1}{|\setI| - 1} .
\end{align}
In practice, we do not know $r$ but want to estimate it from $\tilde{r}$.
First, the probability $p$ can be estimated from the sampled rank by:
\begin{align}
    \hat{p} = \frac{ \tilde{\rank}(i | c) -1}{m}
\end{align}
And thus, an estimator of the rank on the full catalog is
\begin{align}
    \hat{\rank}(i|c) = 1 + \frac{ \tilde{\rank}(i | c) -1}{m} (|\setI|-1) .
\end{align}
The estimated rank $\hat{\rank}$ is a better estimator of the true rank $\rank$ than the sampled rank $\tilde\rank$.
In Section~\ref{sec:experiments}, we compare batch training of WARP and \lambdarank with the sampled rank and the estimated rank.

\subsection{Discussion}
An alternative to estimating the rank, $\hat{\rank}$, would be to estimate $\hat{\alpha}$ from the sampled ranks.
This could follow the ideas of \cite{krichene:sampled_metrics} for estimating a corrected metric from sampled ranks.
For example, for \lambdarank, $\alpha$ is a function of the $\ndcg$ and \cite{krichene:sampled_metrics} suggested methods for estimating the $\ndcg$ metric from sampled ranks $\tilde{\rank}$.
This is a promising direction for future work.

\section{Evaluation}
\label{sec:experiments}

This section investigates the batched versions of \lambdarank and WARP as well as the importance of improved rank estimation.

\subsection{Setup}

Our study is carried out on two well established item recommendation benchmarks: Movielens 20M (ML20M)~\cite{harper:movielens} and Million Song Dataset (MSD)~\cite{bertinmahieux:msd}.
We follow the data preprocessing and data splits from~\cite{liang:vae}\footnote{\url{https://github.com/dawenl/vae_cf}} which have been used by several studies, e.g.,~\cite{shenbin:recvae,kim:hvamp,khawar:swdae,lobel:ract,steck:ease}.
We refer the reader to~\cite{liang:vae} for details about data preprocessing.
The preprocessed datasets contain user interactions with items from catalogs of about 20k movies (ML20M) and 41k songs (MSD).
The dataset split of~\cite{liang:vae} holds out 20\% of the interactions of 20k users (ML20M) and 100k users (MSD) for evaluation purposes.
We train the models on the remaining 80\% of the interactions of the evaluation users and on all the interactions of the remaining users.
We measure the $\recall@20$, $\recall@50$ and $\ndcg@100$ metrics on the evaluation users.
The evaluation users are divided into two sets of equal size: the tuning users and the test users.
We use the average metrics of the tuning users for hyperparameter tuning and we report the average metrics on the test users.
The test set for which we report our results is identical to~\cite{liang:vae}.
Our evaluation protocol differs slightly from~\cite{liang:vae} because we learn the user embeddings of test users together with all remaining users.
The reason is that (i)~unlike autoencoder models (e.g., \cite{liang:vae,steck:ease}) that have only item embeddings, our model also requires user embeddings and (ii)~unlike the iALS training method~\cite{hu:ials}, our gradient descent based training does not offer a closed form equation to compute user embeddings.

Our proposed learning method is general and not tied to a particular model.
In our study, we demonstrate the learning algorithm on the matrix factorization model.

\begin{table}[t]
    \centering
    \addtolength{\tabcolsep}{-0.2em}
    \begin{tabular}{lrrrrrrrr}
       \toprule
& \multicolumn{3}{c}{ML20M} && \multicolumn{3}{c}{MSD} \\
& R20 & R50 & N100 && R20 & R50 & N100\\
\midrule
 WARP (from \cite{shenbin:recvae})  & 0.314 & 0.466 & 0.341 && 0.206 & 0.302 & 0.249  \\
 WARP (ours)                        & 0.345 & 0.480 & 0.382 && 0.240 & 0.350 & 0.292 \\
 \midrule
 LambdaNet (from \cite{shenbin:recvae})    & 0.395 & 0.534 & 0.427 && 0.259 & 0.355 & 0.308  \\
 \lambdarank (ours)                 &   0.379 & 0.503 & 0.401 && 0.269 & 0.370 & 0.314 \\
\bottomrule
    \end{tabular}
    \caption{Comparison of our batched algorithms with sampling to existing results for WARP and \lambdarank.}
    \label{tab:tbl_previous}
\end{table}

\begin{figure*}[t]
    \centering
    \includegraphics[width=0.9\textwidth]{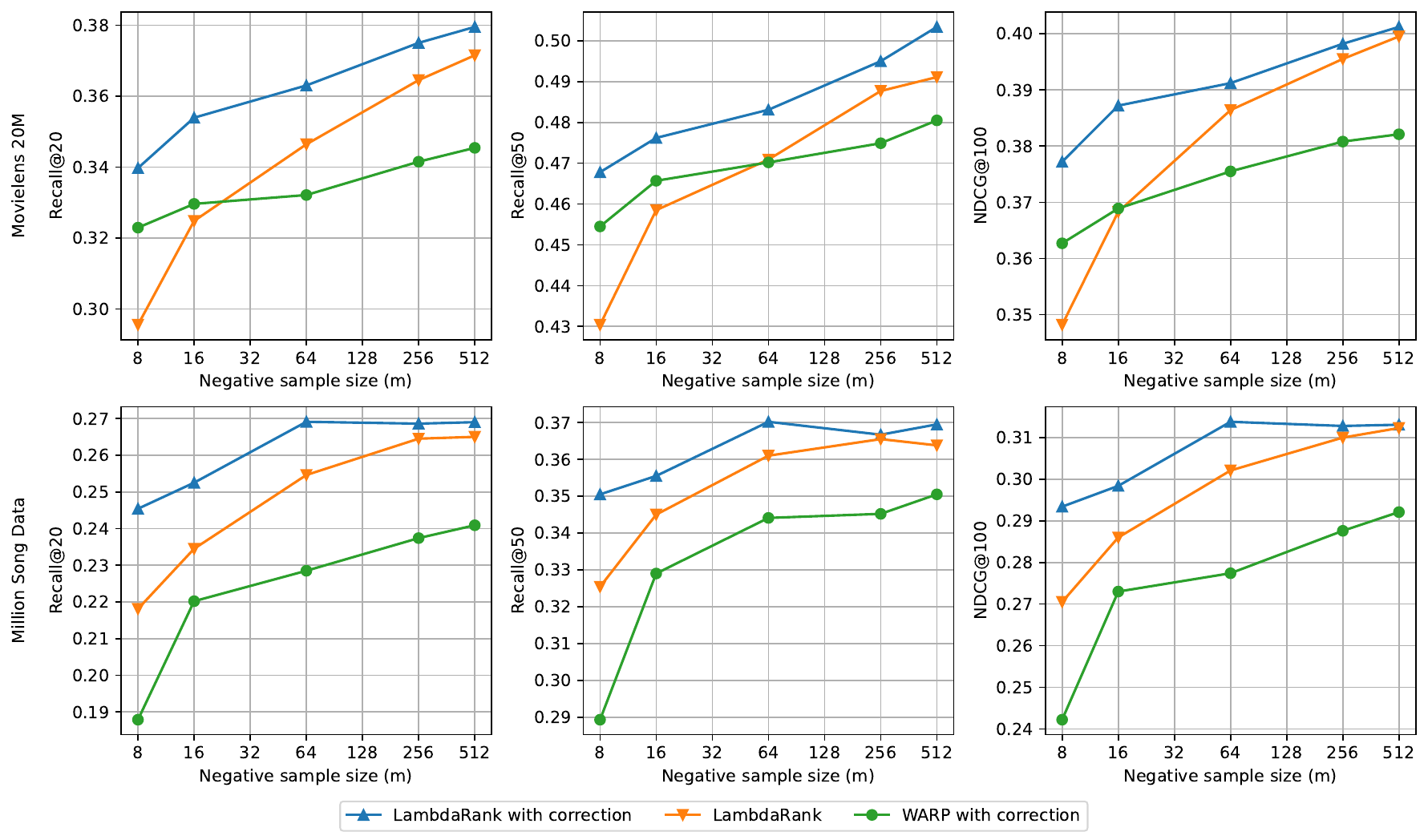}
    \caption{Evaluation of batched versions of WARP and \lambdarank with different negative sample sizes.}
    \label{fig:results}
\end{figure*}

\begin{figure}[ht]
    \centering
    \includegraphics[width=0.7\linewidth]{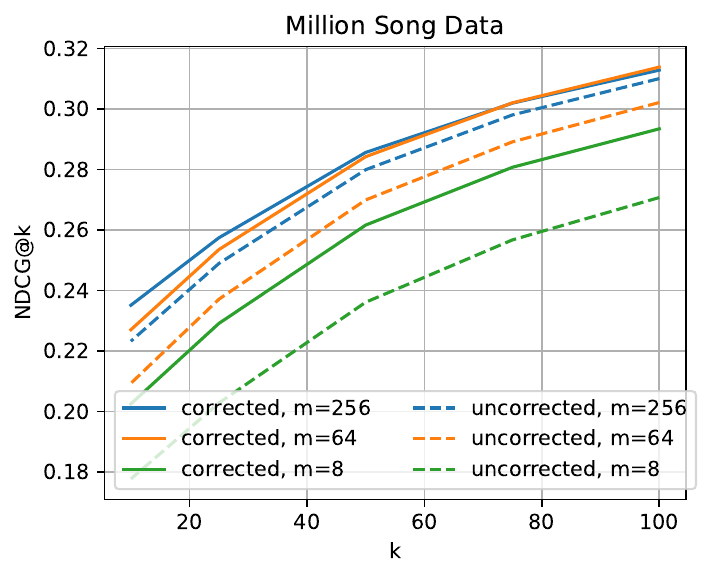}
    \caption{Effect of correction at various thresholds k of $\ndcg@k$ for \lambdarank.}
    \label{fig:eval_thresholds}
\end{figure}

\subsection{Results}

\subsubsection{Batch Training with Negative Sampling}

Table~\ref{tab:tbl_previous} compares our batched algorithms with negative sampling to existing results for WARP and \lambdarank from~\cite{shenbin:recvae}.
Our batch trained WARP with negative sampling improves over previous results on both the MSD and ML20M datasets.
Our \lambdarank with negative sampling outperforms the previously reported \lambdarank results on MSD but underperforms previous results on ML20M.

\subsubsection{Effect of Sample Sizes}

Figure~\ref{fig:results} compares the prediction quality of batch training methods with different negative sample sizes.
The ranks are estimated from the sampled items $\tilde{\setI}$.
For \lambdarank, we show versions with and without correction and for WARP we always apply the correction.
The plots show that increasing the sample size improves the quality.

\lambdarank with correction performs in general better than \lambdarank without correction.
The benefits of correction is particularly strong for small sample sizes.
This finding is consistent with the theory where the bias decreases with larger sample sizes.

In general, \lambdarank (with and without correction) performs better than WARP.
Only for very small sample sizes, WARP performs better than uncorrected \lambdarank.
This indicates that in the small sample regime correction is especially important.

\subsubsection{Effect of Evaluation Threshold}
Figure~\ref{fig:eval_thresholds} shows the quality for different \lambdarank version when varying the cutoff $k$ of the evaluation metric $\ndcg@k$.
In general, the prediction quality of all sample sizes reduces when the metric gets more top-heavy (i.e., $k$ smaller).
While the overall trends look similar, the quality gap between larger sample sizes ($m=256$) and smaller sample sizes ($m=64$) increases when the task gets harder (more top-heavy).

\section{Conclusion}
In this work, we have shown that methods that have been previously proposed for correcting sampled metrics can be applied to improve training algorithms with negative sampling.
We demonstrated this on WARP and \lambdarank, two well-known learning algorithms that depend on the rank of items.
We proposed sampled batch training versions of these algorithms and applied the correction methods to estimate ranks from samples of negative items.
We studied the benefits from the corrected rank estimates empirically.
Our experiments show that the proposed rank estimates improve the prediction quality over the naive computation of ranks on the sample.
Finally, we make a note that the proposed corrections are not restricted to WARP and \lambdarank but can be applied to any algorithm that performs loss calculation using ranking metrics.
A direction for further improvements would be to apply the metric correction methods from~\cite{krichene:sampled_metrics}, in particular trading off the bias-variance when estimating the weighting function $\alpha$ from samples.
This could further improve over correcting the rank as described in our work because the weighting function $\alpha$ is the quantity that we are ultimately interested in.

\bibliographystyle{ACM-Reference-Format}
\balance
\bibliography{paper}

\end{document}